\documentclass[12pt]{iopart}

\begin{document}

\title{Geometric models of particles - the missing ingredient}
\author{Mark J Hadley}
\address{Department of Physics, University of Warwick, Coventry
CV4~7AL, UK\\ email: Mark.Hadley@warwick.ac.uk}

\begin{abstract}
Einstein's dream of describing elementary particles as solutions of a classical field theory is severely limited by our current understanding of Nature. Quantum theory is inconsistent with any local realistic theory such as evolving topological structures in space. It is shown that allowing time reversal as an intrinsic part of the structure of elementary particles allows general relativity to explain quantum theory in principle. Such structures may also explain electric charge and spin-half.
\end{abstract}

\maketitle

\section{Introduction}
Einstein's dream was to find a grand unified theory. For him that meant not only a unification of the forces of nature, but also the unification of particles and fields. He tried repeatedly to find field theories that included gravitation and also had particle like solutions. It is interesting to note that some early attempts were dismissed because they did not have symmetric solutions of unequal masses and opposite charge corresponding to an electron and a proton - the only know elementary particles at the time.

Although the focus of his later work was on classical field theories, Einstein also held the view that a more fundamental realist theory would eventually explain quantum theory. Einstein probed the completeness of quantum theory with his EPR experiment, but it was Bell who used the same thought experiment to show that quantum theory was fundamentally incompatible with any local hidden variable theory. This ruled out any realist explanation compatible with causality.

To model elementary particles with a classical field theory requires solutions:
\begin{enumerate}
    \item with the right properties - such as mass, charge, spin etc
    \item Interactions between them - scattering, creation annihilation etc.
    \item The right behaviour - quantum theory
\end{enumerate}

although Einstein considered the first requirement in detail, the other two were largely ignored. It may have seemed expedient to get solutions first and then examine interactions and mechanics afterwards. However, the second and third requirements place severe constraints upon acceptable solutions and in so doing offer potentially helpful clues.

\section{Interactions and topology change}

If particles are modeled as topological structures of space, then interactions such as annihilation of particles, particle-antiparticle creation and some scattering experiments will require the topology of a region of spacetime to change. However there are powerful theorems due to Geroch \cite{geroch68} and Tipler \cite{tipler} that constrain topology change in classical general relativity. Topology change in classical general relativity  would need to be a counterexample to Geroch's theorem. The possible counterexamples are analysed in Hadley (1999). In summary the possibilities can be classified as:

\begin{enumerate}
\item Singularities
\item Closed timelike curves
\item A lack of time orientation
\end{enumerate}

The first is a breakdown of general relativity. Since General Relativity is expressed in terms of a 4D spacetime manifold, a singularity is not consistent with a description of spacetime as a manifold. Nevertheless some authors (eg Sorkin \cite{sorkin97}) have proposed a reformulation of general relativity to alow singularities where topology change takes place.

For closed timelike curves to appear in a region of space that was previously regular requires negative energy due to Tipler's theorem. While this cannot be ruled out absolutely with current theoretical knowledge it is unwelcome as a postulate.

The third option is that some timelines from the initial surface turn around and return through the same surface. This would be a failure of time-orientability. Further work on structures that lack time orientability have proved fascinating and fruitful.

\section{The Essence of Quantum Theory}

Quantum theory has many strange features, but most are not absolutely unique to quantum theory. For example the uncertainty relationship between position and momentum is common to classical waves, or the wave equations themselves. While some, apparently critical, features are largely irrelevant; such as the inability to make simultaneous measurements of some observables - if this were possible  (even for non-commuting observables) it would not change quantum theory.

A fundamental difference between quantum and classical physics can be found in the logical structure of the propositions. Classical physics satisfies Boolean logic while quantum theory is non-Boolean. The distributive law does not hold, instead the propositions form an orthomodular lattice \cite {beltrametti_cassinelli}. This leads to the requirement to represent probabilities as subspaces of a [complex] vector space rather than as measures of a volume space. It is this fundamental distinction that leads to the dynamical and statistical features of quantum theory.

 Although the relationship between incompatible observables has a non-Boolean characteristic, in any one experiment the propositions satisfy the normal Boolean logic. This is generally expressed by saying that quantum theory is {\em context dependent} - within any single context, the probabilities can be expressed in the normal way.

 A consequence of the logical structure of quantum theory and the way probabilities are represented is an entirely new concept of probability. In all of classical physics probability can be ascribed to our ignorance of some variables (typically the precise initial conditions). In quantum theory this interpretation is not possible, it cannot be described in terms of local hidden variables. As shown by the violations of Bell's inequalities.

The logical structure and the new way to represent probabilities leads inevitably, and uniquely,  to quantum theory and quantum field theory. The continuity and symmetry of spacetime give rise to the usual equations of quantum theory.

\section{The logical structure of acausal spacetimes}
A formal proof that acausal spacetimes lead to the logical structure of quantum theory is given in the paper by Hadley \cite{hadley97}. The proof is formal and can be largely replaced by the statement that an acausal spacetime is context dependent. This can been seen simply in two ways:

With closed timelike curves (CTCs) or a failure of time orientability, it is not in general possible to set up boundary conditions on an {\em initial} surface without some knowledge of {\em future} conditions. Future experiments can set extra boundary conditions that are not redundant. Simple models with Billiards such as those described by Carlini and Novikov \cite{carlini_novikov} provide a good illustration. The standing waves on a string provide a helpful analogy, boundary conditions need to be specified at both ends before the wave can calculated.

An alternative way to look at the context dependence, with the same example, would be to consider the shape of the standing wave as a combination of forward and backward moving waves. A change of time direction at a future experiment would {\em send a signal back} to the start of the experiment. This can be seen as a realisation of the Cramer's transactional interpretation of quantum theory \cite{cramer}.

\section{4-geons}

The original term {\em geon} was coined by Misner and Wheeler \cite{misner_wheeler} to describe a topological structure in space that had particle-like properties. They made no claim that their geons were particles and indeed the analysis above shows that they could never have displayed quantum like properties or interactions. However general relativity is an intrinsically 4 dimensional theory. The space and time components have equal status (the single time component is distinguished only by the signature of the metric).

A 4-geon has non-trivial causal structure as well non-trivial spacial topology. The particle and its evolution are inseparable. Time reversal is an intrinsic part of the measurement process. The 4-geons form the basis for context dependence in classical general relativity. Strictly speaking the non-trivial causal structure also needs some interactions to be taking place. Within the strict limits of classical general relativity, that could most easily be a gravity wave, so that the characteristics would be dependent upon boundary conditions formed by both state preparations and measurement.

\section{Supplementary results}
Manifolds that are not time orientable have some fascinating properties. The integral theorems, linking the enclosed charge to the flux on a closed surface have limited applicability. Stokes theorem requires the enclosed volume to be orientable, while the divergence theorem requires a co-orientation to exist. When the theorems do not apply, it is possible to have the appearance of charge arising from the source free equations because there can be a net surface flux with zero enclosed charge. Sorkin \cite{sorkin77} applied the idea using Stoke's theorem to a geon that was not space orientable and found that it could have net magnetic charge. However a 4-geon that is not time orientable does not have a consistently defined normal vector (it is not co-orientable) and the divergence theorem fails, allowing the appearance of net electric charge. See Diemer and Hadley \cite{diemer_hadley}.

The rotational properties of a non-time orientable manifold are intriguing. It was known that on some 3-manifolds it may not be possible to define a rotational vector field continuously \cite{hendriks}. On an asymptotically flat manifold that did not admit a rotational vector field, it would be possible to define a rotation in the asymptotic region but not possible to extend the rotation throughout the structure. Sorkin \cite{friedman_sorkin} used this result to infer spin-half characteristics in semi-classical gravity. However for a manifold that is not time orientable, in the general case (without additional symmetry) it would not be possible to rotate the asymptotic region and extend it through to the whole structure. The result is described in \cite{hadley2000}. Such a manifold would not have a 360 degree rotation as a symmetry operation, but a 720 degree rotation would be diffeomorphic to the identity. A 4-geon would have the transformation properties of a spinor.

\section{Conclusion}

To the authors knowledge, the first reported solution of general relativity that was not time-orientable, was the famous Einstein-Rosen bridge \cite{einstein_rosen} (when seen as a bridge between two parts of the seme Universe). It is not usually appreciated that a traversable Einstein-Rosen bridge would result in the traveler coming out with the opposite sense of time direction as an external observer. To the outside observer both mouths of the bridge would be black holes but the traveler would come out of one and see it as a white hole, but only because his sense of time is reversed. Probing a time reversing region is not trivial - as described in \cite{hadley2002}. The properties of the bridge arise because the radial co-ordinate becomes the time coordinate at the event horizon - and the wormhole smoothly changes an ingoing radial trajectory into an outward one.

At least in principle, General Relativity with non-trivial causal structure could explain quantum theory and much more besides. It may be the unified theory that Einstein sought for so long.


\section*{References}
\begin{thebibliography}{99}

\bibitem{beltrametti_cassinelli} Beltrametti E G and Cassinelli G 1981
\newblock {\em The Logic of Quantum Mechanics, Vol 15 Encyclopedia of Mathematics and its Applications} Addison Wesley Publishing Company

\bibitem{carlini_novikov} Carlini A and Novikov I D 1996
\newblock {\em Int. J. Mod. Phys.} {\bf D5} 445--480

\bibitem{cramer} Cramer J G 1986
\newblock {\em Rev. Mod. Phys.} {\bf 58} 647--688

\bibitem{diemer_hadley} Diemer T and Hadley M J 1999
\newblock {\em Class. Quantum Grav.} {\bf 16} 3567--3577

\bibitem{einstein_rosen} Einstein A and Rosen N 1935
\newblock{\it Phys. Rev.} {\bf 48} 73--77

\bibitem{friedman_sorkin} Friedman J L and Sorkin R D 1982
\newblock {\em Gen. Rel. Grav.} {\bf 14} 615--620

\bibitem{geroch68} Geroch R 1968
\newblock{\it J. Math. Phys.} {\bf 9} 1739--1744

\bibitem{hadley97} Hadley M J 1997
\newblock{\it Found. Phys. Lett.} {\bf 10} 43--60

\bibitem{hadley98} Hadley M J 1999
{\it Int. J. Theor. Phys.} {\bf 38} 1481--1492
\
\bibitem{hadley2000} Hadley M J 2000
\newblock {\it Clas. Quant. Gravity} {\bf 17} 4187--4194

\bibitem{hadley2002} Hadley M J 2002
\newblock {\it Clas. Quant. Gravity} {\bf 19} 4565-4197

\bibitem{hendriks}Hendriks H 1977
\newblock {\em Bull. Soc. Math.  de France Mem} {\bf 53} 81--96

\bibitem{misner_wheeler} Misner C W and Wheeler J A 1957
\newblock{\it Ann. Phys.} {\bf 2} 525--603

\bibitem{sorkin77} Sorkin R D 1977
\newblock {\em J. Phys.A:Math. Gen.} {\bf 10} 717--725
\bibitem{sorkin97} Sorkin R D 1997
\newblock {\em Int. J. Theor. Phys.} {\bf 36} 2759--2781
\
\bibitem{tipler} Tipler 1976
\newblock {\em Phys. Rev. Lett.} {\bf 37} 879
\



\end{thebibliography}
\end{document}